\documentclass{article}

\usepackage{arxiv}

\usepackage[latin1]{inputenc}
\usepackage[english]{babel}
\usepackage[backend=biber,style=alphabetic,sorting=ynt]{biblatex}
\usepackage{csquotes}
\usepackage{graphicx}
\usepackage{setspace}
\usepackage{xcolor}
\usepackage{booktabs}
\usepackage{amsfonts}       
\usepackage{nicefrac}       
\usepackage{microtype}      
\usepackage{lipsum}

\title{Semantics of Conjectures}
\author{Alessio	Rolfini \\
Alma Mater -- University of Bologna \\
Bologna, Italy \\
\texttt{alessio.rolfini@studio.unibo.it}
}

\begin{document}

\maketitle

\begin{abstract}
This paper aims to expand and detail the notion of formal semantics of  \textit{Conjectures} by applying the theoretic-model approach seen in \cite{HAYES04}, \cite{HAYPSC14} and related works.

After a short introduction to the concepts and basics of \textit{Conjectures}, we will start from the notion of Simple Interpretation of RDF, applying and extending the semantic rules and conditions to fully cover the concepts and features of \textit{Conjectures}.
\end{abstract}

\section{Introduction} \label{Conjectures}
The semantics of RDF makes it impossible to express contradictory points of view that act on the same data.

Every approach applied in the past has not solved the main issue, that is, being able to express different statements whose truth value is not known, or even in contrast with each other, without fully asserting them, therefore making them undoubtedly true.

In this paper we introduce and examine the characteristics of \textit{conjectures}, a proposed extension to RDF 1.1 that by design allows the expressions of graphs whose truth value is unknown.

The approach revolves around two main concepts:
\begin{itemize}
    \item \textit{conjecture}: a \textit{concept whose truth value is not available} and its representation;
    \item \textit{collapse to reality}: a mechanism to fully assert, when the time comes, the truth value of conjectures in their RDF-esque form.
\end{itemize}

Conjectures encapsulate plain RDF statements or other conjectures in specially marked named graphs.
They can be expressed in a strong form:

\begin{verbatim}
CONJECTURE :deVereWroteHamlet {
    :Hamlet dc:creator :EdwardDeVere .
}
\end{verbatim}

or, allowing us to be able to include them in plain RDF 1.1 datasets, in a weak form, where the predicate is expressed as a \textit{conjectural predicate}:

\begin{verbatim}
@prefix conj0001: <http://example.org/exampleDoc#deVereWroteHamlet>

GRAPH :deVereWroteHamlet {
  :Hamlet conj0001:creator :EdwardDeVere .
  conj0001:creator conj:isAConjecturalFormOf dc:creator . 
}
:deVereWroteHamlet prov:wasAttributedTo :JThomasLooney .

\end{verbatim}

Additional statements adding information to the conjecture, but external to it, are the \textit{grounds} or \textit{ground statements}. One of them is the last triple in the last example. 

This paper is organized as follows: 
in section 2 we show a number of fundamental concepts and results of the conceptual model of conjectures. 

In section 3 we introduce the impact of conjectures onto the standard RDF Simple interpretation according to \cite{HAYES04}. 

Sections 4 through 7 describe the simple interpretation of the main concepts of Conjectures: conjectures of individual triples, collapse to reality of individual triples, conjectures and collapses of graphs, and conjectures involving blank nodes. 

In section 8 we discuss the simple interpretation of the additional features of conjectures: nested conjectures, conjectural collapses, ad cascading collapses. 

In section 9 we draw some conclusions about the formal model presented here.

\section{Set-theoretical formal model}

Assume the disjoint sets $\mathcal{I}$ (all IRIs), $\mathcal{B}$ (blank nodes), and $\mathcal{L}$ (literals). 
An RDF triple is a tuple $(s, p, o) \in \mathcal{T} = (\mathcal{I} \cup \mathcal{B}) \times \mathcal{I} \times (\mathcal{I} \cup \mathcal{B} \cup \mathcal{L})$. 

For every RDF predicate $p$, let $\mathcal{S}_{p} \subseteq (\mathcal{I} \cup \mathcal{B})$ be its domain and
$\mathcal{O}_{p} \subseteq (\mathcal{I} \cup  \mathcal{B} \cup \mathcal{L})$ its range. 

We denote with 
\begin{verbatim}
    x { s p o } 
\end{verbatim}

a triple $(s, p, o)$ that is referred to in the examples by
the name $x$.

\textbf{Conjecturing}: Conjecturing is the function $conj : \mathcal{T} \rightarrow \mathcal{T}$ such that, for every RDF triple $t_{1} = (s_{1} , p, o_{1}), conj (t_{1}) = (s, p_{s,o} , o)$ where:
\begin{enumerate}
    \item Identity of subject: $s_{1} = s$.
    \item Identity of object: $o_{1} = o$.
    \item Disjointness: $\forall s_{j}, o_{k}$ such that $(s_{j}, p_{s,o} , o_{k}) \in \mathcal{T}$, we have that $s = s_{j}$ and $o = o_{k}$ . 
\end{enumerate}

\textbf{Conjectures, conjectural predicates, conjecturing triple}: given the triple $(s, p, o) \in \mathcal{T}$, its conjecture is the triple

\[conj((s, p, o)) = (s, p_{s,o}, o) .\] 

Conjectural predicates (or \textit{weak predicates}) of predicate $p$ are all the predicates that are
members of the set $\mathcal{C}onj_{p}$ , such that:

\[ \mathcal{C}onj_{p} = \{cp \in \mathcal{I} | \exists s \in S_{p}, o \in O_{p}, conj((s, p, o)) = (s, cp, o) \}\].

\textbf{Theorem 1}: every conjectural predicate is used in one triple only.

\textbf{Proof}: derives from item 3 of definition 1 (Disjointness):
Given two triples $(s_{1}, p_{s,o}, o_{1}) , (s_{2}, p_{s,o}, o_{2}) \in \mathcal{T}$.

For item 3 of definition 1 (Disjointness), we have that  $s_{1} = s_{2}$ and $o_{1} = o_{2}. \forall (s, p, o) \in \mathcal{T}, \exists! p_{s,o}$ such that $conj((s, p, o)) = (s, p_{s,o} , o)$.

\textbf{Corollary 1}: the function $conj$ is unique (barring predicate name changes).

\textbf{Proof}: derives immediately from Theorem 1.

\textbf{Conjectural Form}: predicate $q$ is said to be a \textit{conjectural form} of $p$ if there exists a pair of subjects and objects $s, o$ such that $conj ((s, p, o)) = (s, q, o)$.

\textbf{Collapsing}: Collapsing is a function $collapses: \mathcal{T} \rightarrow \mathcal{T}$ such that, for every RDF triple $t = (s, p, o)$,
$collapses(s, p, o) = (s, p_{s,o} , o)$ iff $conj((s, p, o)) = (s, p_{s,o} , o)$, and is undefined otherwise.

\textbf{Collapsed predicate; collapse}: let $(s, p_{s,o}, o)$ be a conjecture. 
The collapsed predicate of $p_{s,o}$ is the predicate $p$ such that 
\[collapses (s, p, o) = (s, p_{s,o} , o)\]. 
The collapse is the triple 
\[( (s, p, o), collapse, (s, p_{s,o}, o))\].

\section{RDF Simple interpretation and Conjectures}

In RDF, a simple interpretation $I$ of a vocabulary $V$ consists of:
\begin{enumerate}
\item A non-empty set $IR$ of resources, called the domain or universe of $I$.
\item A set $IP$, called the set of properties of $I$.
\item A mapping $IEXT$ from $IP$ into the powerset of $IR \times IR$ i.e. the set of sets
of pairs $< x, y >$ with $x$ and $y$ in $IR$ .
\item A mapping $IS$ from IRIs into $IR$ - in order to map resources and properties
\item A partial mapping $IL$ from literals into $IR$ - in order to map literals
\end{enumerate}

$IEXT(p)$ is the extension of $p$, that is the set of pairs that are the arguments for which the property $p$ is true.

According to \cite{HAYPSC14}, a semantic extension is a set of additional semantic assumptions that gives IRIs additional meanings on the basis of other specifications or conventions.
When this happens, the semantic extension must conform to the  minimal truth conditions already enunciated, extending from them to accommodate the additional meanings.

Therefore, we extend the RDF Simple Interpretation adding a new set of conjectural properties, $IPC$, disjoint from the set of properties $IP$, where the conjectural predicates are created on the fly.

We add a new mapping $IEXTC$ from $IPC$ to the Cartesian product $IR \times IR$. $IEXTC(cp)$ identifies the pair for which the property $cp$ is true.

Because of the Disjointness property of the conjecturing function $conj$ seen in section 1, the pair satisfying the property $cp$ will always be unique.

We need to specify an additional mapping $CONJFORM$ from $IP$ into $IPC$ to map the conjectural forms of the properties. 

Our full simple interpretation of $I$ in RDF with Conjectures is:

\begin{enumerate}
\item A non-empty set $IR$ of resources, called the domain or universe of $I$.
\item A set $IP$, called the set of properties of $I$. 
\item { \color{blue} A set $IPC$, called the set of conjectural properties of $I$. $IPC \cap IP = \emptyset$ }
\item A mapping $IEXT$ from $IP$ into the powerset of $IR \times IR$ i.e. the set of sets of pairs $< x, y >$ with $x, y \in IR$ .
\item { \color{blue} An injective mapping $IEXTC$ from $IPC$ into  $IR \times IR$, in other words the set of pairs $< x, y >$ with $x, y \in IR$. By definition of injective mapping, if $IEXTC(a)=IEXTC(b)$, then $a=b$, that is, any $cp \in IPC$ uniquely applies to a specific pair $<x , y>$.}
\item { \color{blue} A mapping $CONJFORM$ from $IP$ into $IPC$ in order to map the conjectural forms of the properties  }
\item A mapping $IS$ from IRIs into $IR$ - in order to map resources, properties and conjectural properties
\item A partial mapping $IL$ from literals into $IR$ - in order to map literals
\item {\color{blue} A conjecture might be represented as a set of individual statements composing as a whole the conjecture. This set is a \textit{conjectural graph}}

\end{enumerate}

\begin{figure}[htb]
    \centering
    \includegraphics[width=\textwidth]{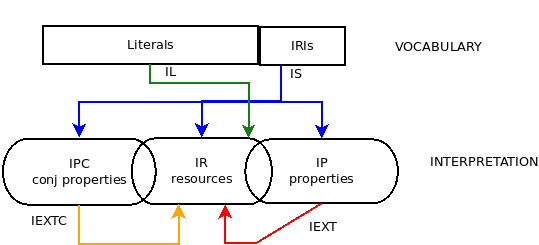}
    \caption{Schematics of the interpretation}
    \label{fig:Interpretation}
\end{figure}

\section{Conjectures}
\begin{itemize}
\item if $E$ is a literal then $I(E) = IL(E)$
\item if $E$ is an IRI then $I(E) = IS(E)$
\item if $E$ is a ground triple $s \: p \: o \: .$ then
    \begin{itemize}
        \item $I(E) = true$ if $I(p) \in IP$ and
        \item the pair $<I(s),I(o)> \in IEXT(I(p))$
        \item otherwise $I(E) = false$.
    \end{itemize}
\item { \color{blue} if $E$ is a Conjecture triple $s \: cp \: o \:.$ then  }
    \begin{itemize}
    \color{blue}
        \item $I(E) = true$ if $I(cp) \in IPC$ and
        \item the pair $<I(s), I(o)> \in IEXTC(I(cp))$
        \item $I(cp) \in CONJFORM(I(p))$ for some $I(p) \in IP$
        \item otherwise $I(E) = false$.
    \end{itemize}
\item if $E$ is a ground RDF graph then $I(E) = false$ if $I(E') = false$ for some triple $E' \in E$, otherwise $I(E) =true$.
\item { \color{blue} if $E$ is a ground conjectural graph then $I(E) = false$ if $I(E') = false$ for some triple $E' \in E$, otherwise $I(E) =true$.}
\end{itemize}

The last clause captures the definition of conjectural graph seen before.

\subsection{Model}
Given a RDF graph $G$, we say that the interpretation $I$ is a model of the graph $G$ if all the triples of graph $G$ are satisfied, that is, if they are true according to rules of $I$.
In this case, we can say that Interpretation $I$ satisfies $G$.

The notion of model is at the basis of the (simple) entailment: 
following standard terminology, we say that $I$ (simply) satisfies $G$ when $I(G)=true$, that $G$ is (simply) satisfiable when a simple interpretation exists which satisfies it, otherwise (simply) unsatisfiable, and that a graph $E$ simply entails a graph $G$ when every interpretation which satisfies $E$ also satisfies $G$. If two graphs $G$ and F each entail the other then they are logically equivalent. \cite{HAYPSC14}

Is our interpretation $I$ a model of this conjecture graph?

\begin{verbatim}
:deVereWroteHamlet { 
   :Hamlet conj001:creator :EdwardDeVere .
   conj001:creator conj:isAConjecturalFormOf dc:creator .
}
\end{verbatim}

\begin{itemize}
    \item $IR=\{dVWH, h, c, cc1, e, iacf\}$
    \item $IP=\{c, iacf\}$
    \item $IPC=\{cc1\}$
\end{itemize}

The functions:

\begin{itemize}
    \item $IS(:deVereWroteHamlet) \rightarrow dVWH$
    \item $IS(:Hamlet)\rightarrow h$
    \item $IS(conj001:creator)\rightarrow cc1$
    \item $IS(:EdwardDeVere)\rightarrow e$
    \item $IS(conj:isAConjecturalFormOf)\rightarrow iacf$
    \item $IS(dc:creator)\rightarrow c$
    \item $IEXT(c)=\emptyset$
    \item $IEXT(iacf)=\{<cc1,c>\}$
    \item $IEXTC(cc1)=\{<h,e>\}$
    \item $CONJFORM(c) = cc1$
\end{itemize}

IEXT(c) is the empty set. We are still dealing with a conjecture, there is no assertion regarding any "real" property (in this case $dc:creator$) yet.

All the clauses seem to hold. We can say that simple interpretation (s-interpretation) $I$ is a model of our graph.

\section{Collapse to reality} \label{collapseToReality}
At a certain point, someone may want to consider our conjectures as true. 

In this case we would need to specify a mechanism in order to express the statements in the conjecture in full force. 

This mechanism is called "collapse to reality".

It consists of exactly two triples added to the base, where:

\begin{itemize}
    \item in the first new triple, the "effective" property is used instead of the corresponding conjectural property.
    \item the first new triple is then declared to "collapse" the conjecture .
\end {itemize}

Two new triples are added, nothing gets replaced or deleted. 
So we can keep track of what's happening and all the relationships between the graphs.

Let's reason on an example of a collapse:

\begin{verbatim}
:attr1 { 
    :Hamlet conj003:creator :Shakespeare .
    conj003:creator conj:isAConjecturalFormOf dc:creator.
}

:attr1Cot {
    :Hamlet dc:creator :Shakespeare .
}

:attr1Cot  conj:collapses :attr1 .

\end{verbatim}

$:attr1$ is collapsed by adding a triple $:attr1Cot$ where the conjectural predicate is replaced by the corresponding "real" predicate.

The final additional triple functions as the explication of the collapse. 
$:attr1Cot$ is declarad to collapse the conjecture $:attr1$ by means of the property "conj:collapses".

For the sake of clarity:
\begin{itemize}
    \item the conjecture triple to be collapsed will be the "conjecture triple";
    \item the new triple collapsing the conjecture will be the "collapsing triple";
    \item the last triple will be the "collapse triple".
\end{itemize}  

\subsection{Interpretation $[I + COLLAPSE]$} \label{I+COLLAPSE}

We need to extend once more our Interpretation with Conjectures $I$ by adding the Collapse to Reality rules.

We define a new mapping $collapses$ from triples to triples that maps the relationship between the conjecture triple and its collapsing triple:

$collapses(s,p,o)=(s,cp,o)$ iff $conj(s,p,o)=(s,cp,o)$.

The collapse collapses the conjecture triple if, and only if, the latter is the (unique) conjecture of the collapsing triple. 

We can clearly see that the "collapses triple" is just the translation into RDF of the definition of collapse we have introduced in section 1, that is, the triple:

\[((s,p,o),\quad collapses,\quad (s,cp,o) )\]  

From a more formal point of view, given a conjectural triple $E$, we add the triple $E_{cot}$ as a collapsing triple $\{s \quad p \quad o .\}$.

The semantic conditions of the collapse to reality should be the following:

\begin{itemize}
\item { \color{blue} 
let $E$ be a conjecture triple
$\\ \{s  \quad  cp  \quad o \quad .\}$

let $E_{cot}$ be a collapsing triple 
$\\ \{ s \quad p \quad o \quad . \}$

and finally let $E_{collapse}$ be the collapse triple
$\\ \{ E_{cot} \quad \mathit{collapses} \quad E \}$

then  } 
    \begin{itemize}
    \color{blue}
        \item $I(E) = true$ if $I(cp) \in IPC$ and
        \item $I(E_{cot}) = true$ if $I(p) \in IP$ and
        \item $CONJFORM(I(p))=I(cp)$ and
        \item the pair $<I(s), I(o)> \in IEXTC(I(cp))$ and
        \item the pair $<I(s), I(o)> \in IEXT(I(p))$ and
        \item $I(E_{collapse}) = true$ if $collapses(I(E_{cot}))=I(E)$, that is:
        $collapses(I(s),I(p),I(o))=(I(s),I(cp),I(o))$ 
        \item otherwise $I(E) = false$ and $I(E_{cot}) = false$ and $I(E_{collapse})=false$.
    \end{itemize}
\end{itemize}

\subsection{Model}

Now, let's check if our Interpretation  $[I + COLLAPSE]$  can be a model of our example graph.
\begin{itemize}
    \item $IR=\{a1, h, cc3, s, iacf, c, a1cot, cl \}$
    \item $IP=\{iacf, c, cl\}$
    \item $IPC=\{cc3\}$
\end{itemize}
The functions:
\begin{itemize}
    \item $IS(:attr1) \rightarrow a1$
    \item $IS(:Hamlet)\rightarrow h$
    \item $IS(conj003:creator)\rightarrow cc3$
    \item $IS(:Shakespeare) \rightarrow s$
    \item $IS(conj:isAConjecturalFormOf)\rightarrow iacf$
    \item $IS(dc:creator)\rightarrow c$
    \item $IS(:attr1Cot)\rightarrow a1cot$
    \item $IS(conj:collapses)\rightarrow cl$
    \item $IEXT(iacf)=\{<cc3,c>\}$
    \item $IEXT(c)=\{<h,s>\}$
    \item $IEXT(cl)=\{<a1,a1cot>\}$
    \item $IEXTC(cc3)=\{<h,s>\}$
    \item $CONJFORM(c) = cc3$
\end{itemize}

Let's check our semantic conditions for the collapse to reality:
\begin{itemize}
    \item $cc3 \in IPC$? Yes; 
    \item $c \in IP$? Yes;
    \item $CONJFORM(c)=cc3$? Yes;
    \item The pair $<h,s> \in IEXTC(cc3)$? Yes;
    \item The pair $<h,s> \in IEXTC(c)$? Yes;
    \item $collapses(h,c,s)=(h,cc3,s)$? Yes, because $CONJFORM(c)=cc3$, hence $conj(h,c,s)=(h,cc3,s)$
\end{itemize}

All of them seem to hold.

Hence, our interpretation $[I + COLLAPSE]$ is a model of our graph.

As we will see in section \ref{CollapseGraphs}, the \textit{collapsing triple} and the \textit{collapse} triple can be safely merged into a \textit{collapse graph}:

\begin{verbatim}
:attr1 { 
    :Hamlet conj003:creator :Shakespeare .
    conj003:creator conj:isAConjecturalFormOf dc:creator.
}

:collapseOfattr1 {
    :Hamlet dc:creator :Shakespeare .
    :collapseOfattr1 conj:collapses :attr1 .
}


\end{verbatim}

\section{Conjectural Graphs and Collapse Graphs}
\subsection{Conjectural Graphs}
A conjectural graph is a representation of a conjecture by means of a set of individual statements composing as a whole the said conjecture.

A ground Conjectural Graph is a graph with no blank nodes.

All triples inside a Conjectural Graph must be either with a conjectural predicate used exactly once, or a "conj:isAConjecturalFormOf" triple connecting a conjectural predicate to an original one.

The semantic conditions for Conjectural Graphs are:
\begin{itemize}
\item if $E$ is a literal then $I(E) = IL(E)$
\item if $E$ is an IRI then $I(E) = IS(E)$
\item { \color{blue} if $E$ is a Conjecture triple $s \: cp \: o \:.$ then  }
    \begin{itemize}
    \color{blue}
        \item $I(E) = true$ if $I(cp) \in IPC$ and
        \item the pair $<I(s), I(o)> \in IEXTC(I(cp))$
        \item $I(cp) \in CONJFORM(I(p))$ for some $I(p) \in IP$ and
        \item $E$ must be unique in the graph
        \item otherwise $I(E) = false$.
    \end{itemize}
\item { \color{blue} if $E$ is a triple $cp conj:isAConjecturalFormOf p$ then }
    \begin{itemize}
    \color{blue}
        \item $I(E) = true$ if $I(cp) \in IPC$ and
        \item $I(p) \in IP$ and
        \item $I(cp) \in CONJFORM(I(p))$ and
        \item $I(conj:isAConjecturalFormOf) \in IP$
        \item the pair $<I(cp), I(p)> \in IEXT(I(conj:isAConjecturalFormOf))$
        \item otherwise $I(E) = false$
    \end{itemize}
\item { \color{blue} if $E$ is a ground conjectural graph then $I(E) = false$ if $I(E') = false$ for some triple $E' \in E$, otherwise $I(E) =true$.}
\end{itemize}

\subsection{Collapse Graphs} \label{CollapseGraphs}

Collapse graphs are graphs that connect explicitly with a conjecture and make it possible to track which conjecture was collapsed.

They must be composed at least of:
\begin{itemize}
    \item the "effective form" of the conjecture to be collapsed 
    \item a "conj:collapse" triple connecting the conjecture to the collapse
\end{itemize}

The semantic conditions are:

\begin{itemize}
\item let $E_{conj}$ be a conjecture triple $s \: cp \: o \: .$
\item let $E_{collapse}$ be a collapse graph
\item if $E$ is a literal then $I(E) = IL(E)$
\item if $E$ is an IRI then $I(E) = IS(E)$
\item { \color{blue} if $E$ is a triple $s \: p \: o \:.$ then  }
    \begin{itemize}
    \color{blue}
        \item $I(E) = true$ if $I(p) \in IP$ and
        \item the pair $<I(s), I(o)> \in IEXT(I(p))$
        \item $CONJFORM(I(p))=I(cp)$ where $cp$ is the conjectural predicate of the conjecture to be collapsed
        \item otherwise $I(E) = false$.
    \end{itemize}
\item { \color{blue} if $E$ is a triple $E_{collapse} \quad conj:collapses \quad E_{conj}$ then }
    \begin{itemize}
    \color{blue}
        \item $I(E) = true$ if $I(conj:collapses) \in IP$ and
        \item the pair $<I(E_{collapse}), I(E_{conj})> \in IEXT(I(conj:collapses))$
        \item otherwise $I(E) = false$
    \end{itemize}
\item { \color{blue} if $E$ is a ground conjectural graph then $I(E) = false$ if $I(E') = false$ for some triple $E' \in E$, otherwise $I(E) =true$.}
\end{itemize}

\section{Blank Nodes}
With conjectures, we would want to be able to express even more "uncertain" concepts involving unnamed entities or unspecified values.

In RDF this is done through \textit{blank nodes}, which indicate the existence of an entity without using a IRI to identify any particular one.

In this section we will use a simple sentence implying the reliance on a blank node, namely  "Muammar al-Qaddafi claimed that the author of Othello was someone who was an Arab" \footnote{He really did it in December 1988 - see "William Shakespeare's Othello" by Jibesh Bhattacharyya, ISBN 9788126904747}. 

In this case we don't know who someone is, and we can't identify it with any IRIs.

Nevertheless, the information we are conveying with our conjecture maintains some degree of meaningfulness, and it definitely is something we could reason upon.

Sticking to our examples' style, we could say it like this: 
"it is conjectured that Othello was written by somebody and this somebody was an Arab. And this claim has been attributed to Muammar al-Qaddafi".

In RDF:
\begin{verbatim}
:ArabWroteOthello { 
  :Othello conj002:creator _:z .
  conj002:creator conj:isAConjecturalFormOf dc:creator .
  _:z dbpedia-owl:nationality :Arab .}

:ArabWroteOthello prov:wasAttributedTo :MalQaddafi .
\end{verbatim}

It comes pretty natural to conceive the term "somebody" as a blank node.

\subsection{Interpretation $[I + A]$}

In order to deal with blank nodes, we need to add a new mapping $A$ from blank nodes into IR.

Therefore, we extend our interpretation $I$:
\begin{itemize}
\item $[I + A](x)=I(x)$ for names
\item $[I + A](x)=A(x)$ if x is a blank node.
\end{itemize}

We add a couple of semantic conditions to our interpretation for blank nodes, one is the "standard" one for RDF graphs, the other one is for conjectures:

\begin{itemize}
    \item if $E$ is a RDF graph then $I(E) = true$ if $[I+A](E) = true$ for some mapping $A$ from the set of blank nodes in $E$ to $IR$, otherwise $I(E) =false$.
    \item { \color{blue} if $E$ is a conjectural graph then $I(E) = true$ if $[I+A](E) = true$ for some mapping $A$ from the set of blank nodes in $E$ to $IR$, otherwise $I(E) =false$.}
\end{itemize}

\subsection{Model}

Is our Interpretation $[I+A]$ with blank nodes a model of our example graph? 

Let's reason on $:ArabWroteOthello$ part only.

Be $A$ our blank nodes mapping into $IR$: $\_:z \rightarrow zz$. 

Our interpretation $[I + A]$ will be: 

\begin{itemize}
    \item $IR=\{awo, o, c, cc2, iacf, n, a, zz \}$
    \item $IP=\{c, iacf, n\}$
    \item $IPC=\{cc2\}$
\end{itemize}

The functions:

\begin{itemize}
    \item $IS(:ArabWroteOthello) \rightarrow awo$
    \item $IS(:Othello)\rightarrow o$
    \item $IS(conj002:creator)\rightarrow cc2$
    \item $IS(dc:creator)\rightarrow c$
    \item $IS(dbpedia-owl:nationality) \rightarrow n$
    \item $IS(:Arab)\rightarrow a$
    \item $IS(conj:isAConjecturalFormOf)\rightarrow iacf$
    \item $IEXT(c)=\emptyset$
    \item $IEXT(iacf)=\{<cc2,c>\}$
    \item $CONJFORM(c) = cc2$
    \item $IEXT(n)= \{<zz,a>\}$
    \item $IEXTC(cc2)=\{<o,zz>\}$
\end{itemize}

Even in this case $IEXT(c)$ is empty because there is no assertion regarding the property $dc:creator$ yet, so our conjecture is, well,  still a conjecture.

The last two functions hold because of the mapping $A$ of our blank node.

All the clauses are true. Our interpretation $I + A$ is a model of our graph.

This approach allows us to reason with the blank nodes in what-if scenarios, where we could define the $A$ mapping from blank nodes to specific resources of our choice, therefore exploring new relationships arising between the triples.

\section{Further applications of conjectures}

\subsection{Conjecture of a conjecture - nested conjectures}

Let's imagine we want to express a conjecture on another conjecture.
Of course, the process can involve as many conjectures over conjectures we want.

Such as:

\begin{verbatim}
:conjecture01 {
    :Hamlet conj004:creator :EdwardDeVere .
    conj004:creator conj:isAConjecturalFormOf dc:creator .
}

:conjecture02 {
 :conjecture01 conj004:wasAttributedTo :JThomasLooney .
 conj004:wasAttributedTo conj:isAConjecturalFormOf prov:wasAttributedTo .
}

:conjecture03 {
:conjecture02 conj004:wasInformedBy <https://bit.ly/3wSH76A> .
conj004:wasInformedBy conj:isAConjecturalFormOf prov:wasInformedBy .
}

:conjecture03 prov:wasAttributedTo :FabioVitali .    
    
\end{verbatim}
They are three different conjectures, one becoming the subject of the other one:
\begin{itemize}
    \item the first one says that Hamlet is conjectured to have been written by Edward De Vere;
    \item the second one says that the previous conjecture could possibly have been made by J. Thomas Looney;
    \item The third one says that it might be that the second conjecture could have been brought to light by the resource with URL https://bit.ly/3wSH76A.
\end{itemize}
The fourth triple says that the last conjecture has been attributed to Fabio Vitali.

Reading it (more or less) backwards:
Fabio Vitali has stated that the resource at https://bit.ly/3wSH76A might have brought to light that J. Thomas Looney could have possibly said that Hamlet is conjectured to have been written by Edward De Vere.

We could imagine it as a stack, or a stair, where at level 0 we have the first conjecture, and then as we get on the higher steps, the conjectures we find are built with the conjecture at the level below as a subject (or object, or both):
\begin{verbatim}
Level 2:            E{3}={E{2}, cp3, o3}
Level 1:                  E{2}=(E{1}, cp2, o2}
Level 0:                        E{1}=(s1, cp1, o1)
\end{verbatim}
 
Developing the Level 2 we see they are nested in each other:
$E{3}=(E{1}, cp2, o2),cp3, o3)=(((s1, cp1, o1), cp2, o2), cp3, o3)$

\subsubsection{Interpretation $[I + NESTEDCONJ]$}

In order to delineate the rules for the conjectures of conjectures (or nested conjectures), we can reason with pairs of conjectures.

As stated before, the conjectures at levels $> 0$ could be based on lower-level conjectures as their subject or object, or both.

For the sake of conciseness, we momentarily limit ourselves to reason with the case of the lower-level conjectures as their subject only.

\begin{enumerate}
    \item if $E_{0}$ is a ground conjecture $(s_{0}, cp_{0}, o_{0})$, $E_{1}= (E_{0}, cp_{1}, o_{1})$
    \item if $E_{1}$ is a ground conjecture $(E_{0}, cp_{1}, o_{1})$, $E_{2}= (E{1}, cp_{2}, o_{2})$
    [...]
    \item if $E_{k-1}$ is a "higher level" ground conjecture $(E_{k-2}, cp_{k-1}, o_{k-1})$, $E_{k}=(E_{k-1}, cp_{k}, o_{k})$  
    \item if $E_{k}$ is a "higher level" ground conjecture $(E_{k-1}, cp_{k}, o_{k})$, $E_{k+1}= (E_{k}, cp_{k+1}, o_{k+1})$  
\end{enumerate}

We should have the following cases:
\begin{itemize}
    \item base case:
        \begin{itemize}
            \item $E_{0}=(s_{0}, p_{0}, o_{0})$
        \end{itemize}
    \item 1st cases:
          \begin{itemize}
              \item $E_{1}=(E_{0}, cp_{1}, o_{1})$
              \item $E_{1}=(s_{1}, cp_{1}, E_{0})$
          \end{itemize}
    \item $k^{th}$ cases:
        \begin{itemize}
            \item $E_{k}=(E_{k-1)}, cp_{k}, o_{k})$
            \item $E_{k}=(s_{k}, cp_{k}, E_{k-1})$
        \end{itemize}
\end{itemize}

Let's extend our interpretation $I$ adding new rules.

The extension will be subdivided into cases, depending on the type of conjectures to be evaluated.

We must also enforce an order on the operations: we start from the conjecture(s) at the "lowest level", that is, the one(s) not involving other conjectures, evaluate them and "climb" up the "stair".

Therefore, the extension to the interpretation $I$ will be:
\newline
\textbf{Base case} - for the lowest level conjecture at level 0:
\begin{itemize}
\item { \color{blue} 
let $E_{0}$ be a conjecture triple
$\\ \{s_{0}  \quad  cp_{0}  \quad o_{0} \quad .\}$
then  } 
    \begin{itemize}
    \color{blue}
        \item $I(E_{0}) = true$ if $I(cp_{0}) \in IPC$ and
        \item the pair $<I(s_{0}), I(o_{0})> \in IEXTC(I(cp_{0}))$
        \item otherwise $I(E_{0}) = false$ 
    \end{itemize}
\end{itemize}

As we "climb" up the "stair" and get to the conjecture $E_{k}$, we can assume that the conjecture $E_{k-1}$ has already been proved by the previous steps.
\newline

\textbf{$K^{th}$ Case S} - for conjectures at the first step and above having another conjecture as the subject:
\begin{itemize}
\item { \color{blue} 
let $E_{k-1}$ be a conjecture triple

let $E_{k}$ be a conjecture triple
$\\ \{ E_{k-1} \quad cp_{k} \quad o_{k}\quad . \}$

then  } 
    \begin{itemize}
    \color{blue}
        \item $I(E_{k}) = true$ if $I(E_{k-1}) = true$ and 
        \item $I(cp_{k}) \in IPC$ and
        \item the pair $<I(E_{k-1}),I(o_{k})> \in IEXTC(I(cp_{k}))$
        \item otherwise $I(E_{k}) = false$
    \end{itemize}
\end{itemize}

\textbf{$K^{th}$ Case O} - for conjectures at the first step and above having another conjecture as the object:
\begin{itemize}
\item { \color{blue} 
let $E_{k-1}$ be a conjecture triple

let $E_{k}$ be a conjecture triple
$\\ \{ s_{k} \quad cp_{k} \quad E_{k-1} \quad . \}$

then  } 
    \begin{itemize}
    \color{blue}
        \item $I(E_{k}) = true$ if $I(E_{k-1}) = true$ and
        \item $I(cp_{k}) \in IPC$ and
        \item the pair $<I(s_{k}), I(E_{k-1})> \in IEXTC(I(cp_{k}))$
        \item otherwise $I(E_{k}) = false$
    \end{itemize}
\end{itemize}

\textbf{$K^{th}$ Case SO} - for conjectures at the first step and above having another conjecture as the subject and yet another one as the object:
\begin{itemize}
\item { \color{blue} 
let $E_{(k-1)a}$ be a conjecture triple

let $E_{(k-1)b}$ be a conjecture triple

let $E_{k}$ be a conjecture triple
$\\ \{ E_{(k-1)a} \quad cp_{k} \quad E_{(k-1)b} \quad . \}$

then  } 
    \begin{itemize}
    \color{blue}
        \item $I(E_{k}) = true$ if $I(E_{(k-1)a}) = true$ and
        \item $I(E_{(k-1)b}) = true$ and
        \item $I(cp_{k}) \in IPC$ and
        \item the pair $<I(E_{(k-1)a}), I(E_{(k-1)b})> \in IEXTC(I(cp_{k}))$
        \item otherwise $I(E_{k}) = false$
    \end{itemize}
\end{itemize}

\subsubsection{Model}
Is our Interpretation $[I + NESTEDCONJ]$ a model of the nested conjectures example seen before?

We define the sets:

\begin{itemize}
    \item $IR=\{c1, h, cc4, edv, iacf, c, c2, cwa4, jtl, pwa, c3, cwib4, http, pwib, fv\}$
    \item $IP=\{c, iacf, pwa, pwib\}$
    \item $IPC=\{cc4, cwa4, cwib4\}$
\end{itemize}

The functions:

\begin{itemize}
    \item $IS(:conjecture01) \rightarrow c1$
    \item $IS(:Hamlet)\rightarrow h$
    \item $IS(conj004:creator)\rightarrow cc4$
    \item $IS(:EdwardDeVere)\rightarrow edv$
    \item $IS(conj:isAConjecturalFormOf)\rightarrow iacf$
    \item $IS(dc:creator)\rightarrow c$
    \item $IS(:conjecture02) \rightarrow c2$
    \item $IS(conj004:wasAttributedTo) \rightarrow cwa4$
    \item $IS(:JThomasLooney)\rightarrow jtl$
    \item $IS(prov:wasAttributedTo)\rightarrow pwa$
    \item $IS(:conjecture03) \rightarrow c3$
    \item $IS(conj004:wasInformedBy) \rightarrow cwib4$
    \item $IL(<https://bit.ly/3wSH76A>) \rightarrow http$
    \item $IS(prov:wasInformedBy)\rightarrow pwib$
    \item $IS(:FabioVitali) \rightarrow fv$
    \item $IEXT(iacf)=\{<cc4,c>,<cwa4,pwa>,<cwib4,pwib>\}$
    \item $IEXT(c)=\emptyset$
    \item $IEXT(pwa)=\{<c3,fv>\}$
    \item $IEXT(pwib)=\emptyset$
    \item $IEXTC(cc4)=\{<h,ev>\}$
    \item $IEXTC(cwa4)=\{<c1,jtl>\}$
    \item $IEXTC(cwib4)=\{<c2,http>\}$
    \item $CONJFORM(c) = cc4$
    \item $CONJFORM(cwa4) = pwa$
    \item $CONJFORM(cwib4) = pwib$
\end{itemize}

Let's check the validity of the rules of the new semantic extension.

We must start from the conjecture not depending on other conjecture, that is $:conjecture01$, and we use the Base Case:

\begin{itemize}
        \item Is $cc4 \in IPC$? Yes
        \item Is the pair $<h, ev> \in IEXTC(cc4)$ Yes
    \end{itemize}

The "base" conjecture seems to hold.
We can say that $c1 = true$.
\newline

Now we "climb the stair" to $:conjecture02$. Since it is based on another conjecture ($:conjecture01$, already proved true) as its subject, we use "$k^{th}$ Case S" 

    \begin{itemize}
        \item is $c1 = true$? Yes
        \item is $cwa4 \in IPC$? Yes
        \item is the pair $<c1,jtl> \in IEXTC(cwa4)$? Yes
    \end{itemize}

Then we can say that $c2 = true$.
\newline

Let's climb one step higher. $:conjecture03$ is based on $:conjecture02$ as its subject. We still use "$k^{th}$ Case S"

    \begin{itemize}
        \item is $c2 = true$? Yes
        \item is $cwib4 \in IPC$? Yes
        \item is the pair $<c2,http> \in IEXTC(cwib4)$? Yes
    \end{itemize}

$\rightarrow c3 = true$.
\newline
The last triple $:conjecture03 prov:wasAttributedTo :FabioVitali$ is satisfied by the rules of  the simple intepretation $I$
\newline
Everything seems to hold.

We can say that our Interpretation $[I + NESTEDCONJ]$ satisfies all the triples of the graph.

\subsection{Conjecture of a collapse} \label{ConjectureOfACollapse}
Let us consider the following sentence: 

\begin{quote}According to Encyclopaedia Britannica (\url{https://bit.ly/3qgakFT}), Samuel Johnson attributed Hamlet to William Shakespeare, and he was right in saying so.\end{quote}

This sentence is more than a mere collapse: it is a conjecture by an article in Encyclopedia Britannica a) attributing to Samuel Johnson a conjecture (about the authorship of Hamlet), and b) expressing total confidence in such conjecture (i.e., collapsing it).  

Its representation is:

\begin{verbatim}
:attribution01 {
  :Hamlet conj005:creator :WilliamShakespeare .
  conj005:creator conj:isAConjecturalFormOf dc:creator .
}

:collapseOfAttribution01 {
  :attribution01 conj005:wasAttributedTo :SamuelJohnson .
  conj005:wasAttributedTo conj:isAConjecturalFormOf prov:wasAttributedTo .

  :collapseOfAttribution01 conj005:collapses :attribution01 .
  conj005:collapses conj:isAConjecturalFormOf conj:collapses .
}

:collapseOfattribution01 prov:wasInformedBy <https://bit.ly/3qgakFT> .
\end{verbatim}

In this case we have a conjecture of a collapse that, if collapsing, as a side effect, triggers the collapse of another conjecture, therefore asserting it in full force.
Please note that, in its current and "uncollapsed" status, everything is conjectured.

We can see the familiar predicate $collapses$ in its conjectural form, not yet in force in this case.
This predicate, once met in its "effective" form, will trigger the collapse of its subject. 
We will see it in the next section.

\subsection{Cascading collapses}

Cascading collapses are all the collapses that take place recursively when multiple and nested conjectures of collapses collapse.

For example, what if we wanted to state that the $collapseOfAttribution01$ is true? We would collapse it by adding the "collapsing triple" and the "collapse triple" as per the rules of Interpretation [I + COLLAPSE] seen before in \ref{I+COLLAPSE}.

For the sake of simplicity, we resort to the notion of collapse graph (\ref{CollapseGraphs}). According to that, we can safely group the added triples into the collapse graph $:collapseOfcollapseOfAttribution01$.

Now that our conjecture of a collapse is declared true, therefore the collapse is true and effective, we have to enforce it, collapsing $:attribution01$

Generalising, once a conjecture of a collapse is declared true, we must enforce a new special rule for its collapse to reality:

Whenever $collapses$ is met in its  effective form inside a collapse graph, its object, if it is a conjecture, must be collapsed.

If the object is a conjecture triple $s \:cp \:o$, we need to add the \textit{collapsing triple} and the \textit{collapse triple}.

Getting back to our example, we can see that the \textit{collapse triple} is already defined in its effective form, in fact we have that:
\begin{verbatim}
    :collapseOfAttribution01 conj:collapses :attribution01 .
\end{verbatim}

So we just need to add the collapsing triple, stating the conjecture in its full force, in the graph where the $conj:collapses$ property is met in full force.

The result will be as follows:

\begin{verbatim}
:attribution01 {
  :Hamlet conj005:creator :WilliamShakespeare .
  conj005:creator conj:isAConjecturalFormOf dc:creator .
}

:collapseOfAttribution01 {
  :attribution01 conj005:wasAttributedTo :SamuelJohnson .
  conj005:wasAttributedTo conj:isAConjecturalFormOf prov:wasAttributedTo .

  :collapseOfAttribution01 conj005:collapses :attribution01  .
  conj005:collapses conj:isAConjecturalFormOf conj:collapses .
}
:collapseOfattribution01 prov:wasInformedBy <https://bit.ly/3qgakFT> .

:collapseOfcollapseOfAttribution01  {
    :attribution01 prov:wasAttributedTo :SamuelJohnson .
    
    :collapseOfAttribution01 conj:collapses :attribution01 .

    :collapseOfcollapseOfAttribution01 conj:collapses :collapseOfAttribution01 . 
       .

    :Hamlet dc:creator :WilliamShakespeare.
}

\end{verbatim}

A collapse graph is built for the first collapse of the conjecture of the collapse, and then, at the end, the final "effective" triple.

We need to extend the Interpretation [I + COLLAPSE] seen before in \ref{I+COLLAPSE}.

After we collapse the conjecture of the collapse following the rules in [I+COLLAPSE], we will find the property $collapses$ in its effective form inside the collapse graph.

We need to extend the notion of $collapses$ defining it from collapse graphs to conjectural graphs:
$collapses(G)=(E)$ iff $\forall (s,cp,o) \in E, conj(s,p,o)=(s,cp,o) with (s,p,o) \in G$.

If the object of the $collapses$ property is a triple, which can be considered as a special case of a graph, we will add its effective form to the conjectural graph $collapses$ is in. 

If the object of $collapses$ property is a conjectural graph, we need to add the effective forms of all its conjectures.

One important thing to be aware of is that we disregard the subject of the $collapses$ property: the effective form of the conjecture(s) in the object(s) is added to the \textit{current} collapse graph no matter their subject(s).

In the semantic conditions below we will simplify for clarity by adopting the notion of a generic graph as the theoretical subject of a $collapses$ property. The graphs will in fact be more than one, if the subjects of the conjectures which are objects of $collapses$ are distinct (as in the example above). 
Also The conjecture graphs/triples can be more than one.

The semantic conditions are:

\textbf{conjecture triple:}
\begin{itemize}
\item { \color{blue} 
let $E$ be a conjecture triple
$\\ \{s  \quad  cp  \quad o \quad .\}$

let $E_{g}$ be a generic graph

let $E_{cg}$ be a collapse graph 
$\\ \{ E_{g} \quad \mathit{conj:collapses} \quad E \\
       s \quad p \quad o . \}$

then  } 
    \begin{itemize}
    \color{blue}
        \item $I(E) = true$ if $I(cp) \in IPC$ and
        \item $I(E_{cg}) = true$ if $I(p) \in IP$ and
        \item $CONJFORM(I(p))=I(cp)$ and
        \item the pair $<I(s),I(o)> \in IEXTC(I(cp))$ and
        \item the pair $<I(s),I(o)> \in IEXT(I(p))$ and
        \item $I(conj:collapses) \in IP$ and
        \item the pair $<I(E_{g}),I(E> \in IEXT(I(conj:collapses))$
        \item otherwise $I(E) = false$ and $I(E_{cg})$ is false.
    \end{itemize}
\end{itemize}

\textbf{conjectural graph:}
If we are dealing with a conjectural graph as the object of $conj:collapses$:
\begin{itemize}
\item { \color{blue} 
let $E$ be a conjectural graph
$\\ \{s_{0}  \quad  cp_{0}  \quad o_{0} \quad . \\
      {[...]} \\
      s_{k}  \quad  cp_{k}  \quad o_{k} \quad . \}$

let $E_{g}$ be a generic graph

let $E_{cg}$ be a collapse graph 
$\\ \{ s_{0} \quad p_{0} \quad o_{0} . \\
       {[...]} \\
       s_{k}  \quad  p_{k}  \quad o_{k} \quad . \\
       E_{g} \quad \mathit{conj:collapses} \quad E\}$

then  } 
    \begin{itemize}
    \color{blue}
        \item $I(E) = true$ if $I(cp_{k}) \in IPC \forall cp_{k} in E$ and
        \item $I(E_{cg}) = true$ if $I(p_{k}) \in IP \forall p_{k} in E_{cg}$ and
        \item $CONJFORM(I(p_{k}))=I(cp_{k}) \forall p_{k} in E$ and
        \item any pair $<I(s_{k}),I(o_{k})> \in IEXTC(I(cp_{k}))$ and
        \item any pair $<I(s_{k}),I(o_{k})> \in IEXT(I(p_{k}))$ and
        \item $I(conj:collapses) \in IP$ and
        \item the pair $<I(E_{g}),I(E)> \in IEXT(I(conj:collapses))$
        \item otherwise $I(E) = false$ and $I(E_{cg})$ is false.
    \end{itemize}
\end{itemize}

\subsubsection{Model}

Let's check once again if the interpretation can be a model for our last example.

We define the sets:

\begin{itemize}
    \item $IR=\{a1, h, cc5, ws, iacf, c, coa1, c5wat, sj, pwat, c5cl, cl, pwib, url, ccoa1  \}$
    \item $IP=\{c, iacf, pwat, pwib, cl\}$
    \item $IPC=\{cc5, c5wat, c5cl\}$
\end{itemize}

The functions:

\begin{itemize}
    \item $IS(:attribution01) \rightarrow a1$
    \item $IS(:Hamlet)\rightarrow h$
    \item $IS(conj005:creator)\rightarrow cc5$
    \item $IS(:WilliamShakespeare)\rightarrow ws$
    \item $IS(conj:isAConjecturalFormOf)\rightarrow iacf$
    \item $IS(dc:creator)\rightarrow c$
    \item $IS(:collapseOfAttribution01) \rightarrow coa1$
    \item $IS(conj005:wasAttributedTo) \rightarrow c5wat$
    \item $IS(:SamuelJohnson)\rightarrow sj$
    \item $IS(prov:wasAttributedTo)\rightarrow pwat$
    \item $IS(conj005:collapses)\rightarrow c5cl$
    \item $IS(conj:collapses)\rightarrow cl$
    \item $IS(prov:wasInformedBy)\rightarrow pwib$
    \item $IL(<https://bit.ly/3qgakFT>)\rightarrow url$
    \item $IS(:collapseOfcollapseOfAttribution01) \rightarrow ccoa1$
    \item $IEXT(c)=\{<h, ws>\}$
    \item $IEXT(iacf)=\{<cc5,c>, <c5wat,pwat>, <c5cl,cl>\}$
    \item $IEXT(pwat)=\{<a1, sj>\}$
    \item $IEXT(cl)=\{<a1, coa1>, <coa1, ccoa1>\}$
    \item $IEXT(pwib)=\{<coa1, url>\}$
    \item $IEXTC(cc5)=\{<h,ws>\}$
    \item $IEXTC(c5wat)=\{<a1,sj>\}$
    \item $IEXTC(c5cl)=\{<a1,coa1>\}$
    \item $CONJFORM(c) = cc5$
    \item $CONJFORM(pwat) = c5wat$
    \item $CONJFORM(cl) = c5cl$
\end{itemize}

Let's check now the rules. 
We focus the check on the cascading collapses only. 

We need to proceed in steps. The first $conj:collapses$ in our final collapse graph has $:attribution01$ as its object.
We can consider $:attribution01$ as a single triple, as it contains only one conjecture:

\begin{itemize}
\color{blue}
    \item is $cc5 \in IPC$? Yes;
    \item is $c \in IP$? Yes;
    \item is $CONJFORM(c)=cc5$? Yes;
    \item is the pair $<h,ws> \in IEXTC(cc5)$? Yes;
    \item is the pair $<h,ws> \in IEXT(c)$? Yes;
    \item is $cl \in IP$? Yes;
    \item the pair $<a1,coa1> \in IEXT(cl)$?Yes.
\end{itemize}

It seems to hold.

Let's get to the second $con:collapses$ of our final collapse graph. It has $:collapseOfAttribution01$ as its object.
It is a graph. Let's apply rules for the graphs:

    \begin{itemize}
    \color{blue}
        \item $c5wat, c5cl \in IPC$? Yes;
        \item $pwat, cl \in IP$? Yes;
        \item $CONJFORM(c5wat)=pwat; CONJFORM(c5cl)=cl?$ Yes;
        \item $<a1,sj> \in IEXTC(c5wat)$? Yes; $<a1, coa1> \in IEXTC(c5cl)$? Yes;
        \item $<a1,sj)> \in IEXT(pwat)$? Yes; $<a1, coa1> \in IEXT(cl)$? Yes;
        \item $cl \in IP$? Yes;
        \item the pair $<a1,coa1)> \in IEXT(cl)$? Yes
    \end{itemize}

It can be easily verified that all the other triples of the example can be satisfied by the interpretation [I + COLLAPSE].

Our interpretation is a model of the graph.

\section{Conclusions}
In this paper we have attempted to build a formal representation of the semantics of the \textit{conjectures}.

We have started with the formal model and extended the simple interpretation and model of RDF 1.1.

Then we have explored the key features of the \textit{conjectures} and verified that they fit in the semantics, expanding it when needed.

The \textit{conjectures} allow to express concepts whose truth value is unknown, without asserting it, something that is currently missing in RDF.

With this work we have demonstrated that, with a somewhat limited extension to its model, it is possibile to add this powerful feature to RDF 1.1.

\printbibliography[
title={Bibliography}
]
\end{document}